\documentclass[11pt]{article}
\usepackage{amsmath}
\usepackage{amssymb}
\usepackage{bbm}
\usepackage{epsfig}
\allowdisplaybreaks[4]
\begin{document}
\title{Stability of the tree-level vacuum in two Higgs doublet models
against charge or CP spontaneous violation}

\author{P.M. Ferreira$^{1,}$~\footnote{ferreira@cii.fc.ul.pt}, 
R. Santos$^{1,2,}$~\footnote{rsantos@cii.fc.ul.pt} and 
A. Barroso$^{1,}$~\footnote{barroso@cii.fc.ul.pt}\\
$^1$ Centro de F\'{\i}sica Te\'orica e Computacional, \\
Universidade de Lisboa, Av. Prof. Gama Pinto, 2, 1649-003 Lisboa, Portugal \\
$^2$ Instituto Superior de Transportes e Comunica\c{c}\~oes,\\
Campus Universit\'ario R. D. Afonso Henriques, 2330-519 Entroncamento, Portugal
}
\date{June, 2004} 
\maketitle
\noindent
{\bf Abstract.} We show that in two Higgs doublet models at tree-level the 
potential minimum preserving electric charge and CP symmetries, when it exists, 
is the global one. Furthermore, we derived a very simple condition, involving 
only the coefficients of the quartic terms of the potential, that guarantees 
spontaneous CP breaking.  
\vspace{1cm}

In the Standard Model (SM) of the electroweak interactions the existence of the 
scalar Higgs doublet is a fundamental piece of the theory. Through it $SU(2)_W 
\times U(1)_Y$ gauge invariance is broken, the $W^\pm$ and $Z^0$ bosons and the
fermions acquire their masses and the renormalisability of the theory is 
preserved. Despite its importance, the scalar sector of the SM has not yet been 
directly tested and there is considerable interest in studying its extensions. 
The simplest of those extensions is the two Higgs doublet model (2HDM).
One of the main reasons of interest in this class of models is the possibility 
of having spontaneous CP violation~\cite{lee}, thus helping to 
solve the baryogenesis problem~\cite{err} (for a review, see~\cite{sher}). One
problem of these models, though, is their immense parameter space - the most 
general potential that preserves the SM gauge group and does not explicitly 
break CP has ten independent parameters, and little is known of their allowed
values. A similar difficulty afflicts Supersymmetric models, where the parameter
space is generally larger. One idea that has been applied to Supersymmetric 
theories to restrict their allowed parameter space is to use charge and colour 
breaking (CCB) bounds. If a given combination of parameters causes the 
appearance in the potential of a minimum where charged/coloured fields have 
vacuum expectation values (vevs), then that combination should be rejected. This
appealing idea was introduced by Fr\'ere {\em et al}~\cite{fre} and applied, in 
numerous papers, to several supersymmetric theories~\cite{eve}. Phenomenological
analysis of supersymmetric Higgs masses use this tool to increase the models' 
predictive power~\cite{phe}. It is therefore of interest to apply similar 
techniques to the 2HDM and try to limit its parameter space. The scalars of this
theory have no colour quantum numbers but there are charged fields so charge 
breaking (CB) extrema are in principle possible. Recent work in 
2HDM~\cite{haber}, for instance, assumed that the choice of parameters made was 
such that the scalar potential respected the $U(1)_{em}$ gauge symmetry.

In reference~\cite{vel} it was shown that to make sure that there were no 
stationary points corresponding to charge or CP spontaneous breaking, one had to
restrict the parameter space to eight independent parameters. This leads to two
independent potentials, stable under renormalisation because they are protected 
by a $Z_2$ or $U(1)$ global symmetries. It is interesting to stress that these 
are the usual symmetries introduced to prevent flavour changing neutral 
currents. In this letter we go back to the study of the tree-level vacuum of the
most general 2HDM potential without explicit CP violation, with ten parameters. 
We prove that for this potential the minimum that preserves CP and $U(1)_{em}$ 
(from now on named the ``Normal minimum"), when it exists, is {\em global}. This
is a very powerful result, in that it assures that the Normal vacuum is stable,
it cannot tunnel to other minima. This letter is structured as follows: we will 
review the model and the Normal minimum in section~\ref{sec:sm}, then proceed to
look at the possibility of charge breaking stationary points, proving there can 
be no CB minima if the potential has a Normal minimum. Likewise, in 
section~\ref{sec:cp}, we will prove the equivalent result for spontaneous CP 
breaking stationary points. In section~\ref{sec:mass} we will look at the mass 
matrices of the three types of stationary points. 

\section{The Normal minimum}
\label{sec:sm}

We will be working with the most general two Higgs doublet model invariant under
the gauge group $SU(2)_W \times U(1)_Y$ that does not explicitly break CP, 
following the conventions of reference~\cite{vel}. This model is built with two 
scalar doublets of hypercharge $Y=1$, 
\begin{equation}
\Phi_1 = \begin{pmatrix} \varphi_1 + i \varphi_2 \\ \varphi_5 + i \varphi_7
\end{pmatrix} \;\; , \;\; \Phi_2 = \begin{pmatrix} \varphi_3 + i \varphi_4 \\ 
\varphi_6 + i \varphi_8 \end{pmatrix} \;\; .
\end{equation}
This numbering of the $\varphi$ fields might seem odd, but it is the most 
convenient for the latter calculation of the mass matrices. As was shown 
in~\cite{vel} this potential has ten independent parameters and may be written 
as
\begin{align}
V \;\;=& a_1\, x_1\, + \,a_2\, x_2\, + \,a_3 x_3 \,+\nonumber \\
  & b_{11}\, x_1^2\, +\, b_{22}\, x_2^2\, +\, b_{33}\, x_3^2\,
 +\, b_{44}\, x_4^2\, +\, b_{12}\, x_1 x_2\, +\, b_{13}\, x_1 x_3\, + \,b_{23}\,
 x_2 x_3 \;\; ,
\label{eq:pot}
\end{align}
with 
\begin{align}
x_1\,\equiv\,|\Phi_1|^2 &= \varphi_1^2 + \varphi_2^2 + \varphi_5^2 + \varphi_7^2
\nonumber \\
x_2\,\equiv\,|\Phi_2|^2 &= \varphi_3^2 + \varphi_4^2 + \varphi_6^2 + \varphi_8^2
\nonumber \\
x_3\,\equiv\,Re(\Phi_1^\dagger\Phi_2) &= \varphi_1 \varphi_3 + \varphi_2 
\varphi_4 + \varphi_5\varphi_6 + \varphi_7\varphi_8 \nonumber \\
x_4 \,\equiv\,Im(\Phi_1^\dagger\Phi_2) &= \varphi_1 \varphi_4 -  \varphi_2 
\varphi_3 + \varphi_5 \varphi_8 - \varphi_6 \varphi_7 \;\;.
\label{eq:x}
\end{align}
The $a_i$ parameters have dimensions of squared mass, the $b_{ij}$ parameters
are dimensionless~\footnote{These parameters have well-known relations to the 
masses and couplings of the physical Higgs particles of the model, see, for
instance, \cite{haber}.}, the fields $\varphi_i$ are real functions. Let us 
introduce a new notation that will be extremely useful: we define a vector $A$ 
and a square symmetric matrix $B$ as
\begin{equation}
A\;=\;\begin{bmatrix} a_1 \\ a_2 \\a_3 \\ 0 \end{bmatrix} \;\;\; , \;\;\; 
B\;=\;\begin{bmatrix} 2 b_{11} & b_{12} & b_{13} & 0 \\ b_{12} & 2 b_{22} & 
b_{23} & 0 \\ b_{13} & b_{23} & 2 b_{33} & 0 \\ 0 & 0 & 0 & 2 b_{44} 
\end{bmatrix} \;\; .
\label{eq:ab}
\end{equation}
Defining the vector $X\, =\, (x_1\,,\,x_2\,,\, x_3\,,\,x_4)$, we can rewrite the
potential~\eqref{eq:pot} in the more compact form
\begin{equation}
V \;=\; A^T\,X \;+\; \frac{1}{2}\,X^T \,B\,X \;\;\; .
\label{eq:vm}
\end{equation}
The Normal minimum corresponds to $\varphi_5 = v_1$, $\varphi_6 = v_2$ and all 
the 
remainder $\varphi_i$ equal to zero. This gives, from the above definitions, 
$x_1 = v_1^2$, $x_2 = v_2^2$, $x_3 = v_1 v_2$ and $x_4 = 0$. We can write the 
relevant minimisation conditions as
\begin{align}
\frac{\partial V}{\partial v_1} = 0 &\Leftrightarrow \frac{\partial V}{\partial 
x_1} \frac{\partial x_1}{\partial v_1} \,+\,\frac{\partial V}{\partial x_3} 
\frac{\partial x_3}{\partial v_1} = 0 \nonumber \\
\frac{\partial V}{\partial v_2} = 0 &\Leftrightarrow \frac{\partial V}{\partial
x_2} \frac{\partial x_2}{\partial v_2} \,+\,\frac{\partial V}{\partial x_3}
\frac{\partial x_3}{\partial v_2} = 0  \;\;\;.
\label{eq:min}
\end{align}
Let us define the vector $V^\prime$, with components $V^\prime_i = \partial V/
\partial x_i$, evaluated at the minimum. From the above it is plain that
\begin{equation}
V^\prime \; = \; \begin{bmatrix} V^\prime_1 \\ V^\prime_2 \\ V^\prime_3 \\ 
V^\prime_4 \end{bmatrix} \; = \; -\frac{V^\prime_3}{2 v_1 v_2}\, \begin{bmatrix}
v_2^2 \\ v_1^2 \\ - 2 v_1 v_2 \\ 0 \end{bmatrix} \;\; . 
\label{eq:vl}
\end{equation}
Looking at the expressions for the two first components of $V^\prime$ it is 
obvious that, regardless of the values of $v_1$, $v_2$ and $V^\prime_3$, 
$V^\prime_1$ and $V^\prime_2$ have the same sign. From this point forward we 
will use $X_{N}$ to designate the vector $X$ evaluated at the minimum, that is, 
with components $(v_1^2\,,\,v_2^2\,,\,v_1\,v_2\,,\, 0)$. In this notation it is 
trivial to realize that $X_{N}^T\,V^\prime \,=\,0$. Direct analysis of the 
potential~\eqref{eq:pot} also shows that we can write $V^\prime$ in matrix form 
as 
\begin{equation}
V^\prime \,=\, A\,+\,B\, X_{N}\;\;\; .
\label{eq:vlm}
\end{equation}
The potential~\eqref{eq:pot} is a sum of quadratic and quartic polynomials,
let us call them $p_2$ and $p_4$; by performing the sum $\sum_i \, v_i\,(
\partial V /\partial v_i)$ it is very simple to show that the minimisation 
conditions imply $2\,p_2 \,+\,4\,p_4 \,=\,0$ at the minimum. As such the value 
of the potential at this stationary point - which we designate by $V_{N}$
- may be written as:
\begin{equation}
V_{N} \;\; = \;\; \frac{1}{2}\,A^T\,X_{N} \;\; = \;\; -\frac{1}{2}\,X^T_{N}\,
B\,X_{N} \;\;.
\label{eq:vmin}
\end{equation}
We have been speaking of the 
Normal minimum but the conditions~\eqref{eq:min} only assure us that
the potential has a stationary point. To ensure we are at a minimum we must
analyse the second derivatives of $V$ - that is, the matrix of the squared 
scalar masses - and reject all combinations of parameters $\{a_i\,,\,b_{jk}\}$ 
for which any of the non-zero eigenvalues are negative (this matrix has three 
zero eigenvalues corresponding to the Goldstone bosons, see 
section~\ref{sec:mass}). In particular we observe that the squared mass of the 
charged Higgs is given by $M^2_{H^\pm}\,=\, V^\prime_1\,+\,V^\prime_2$. So the 
Normal minimum exists if $V^\prime_1\,+\,V^\prime_2 >0$. Since we 
have already shown that $V^\prime_1$ and $V^\prime_2$ have the same sign this 
tells us that both quantities are positive. Therefore we obtain
\begin{equation}
V^\prime_1 \;\; =\;\; \left( -\frac{V^\prime_3}{2 v_1 v_2}\right) \; v_2^2 \; > 
\; 0 \;\; ,
\end{equation}
and so we conclude that {\em the quantity $-V^\prime_3/(2 v_1 v_2)$ is 
positive}. This conclusion will be fundamental later on. 

\section{Charge breaking}
\label{sec:cb}

The potential~\eqref{eq:pot} has only three types of non-trivial stationary 
points~\cite{lee,sher}: the Normal one; a charge-breaking stationary point,
where one of the charged fields $\varphi$ has a non-zero vev; and a CP-breaking
stationary point where one of the imaginary neutral component fields has a 
non-zero vev.  In this section we will deal with charge breaking (CB), 
specifically, the configuration where the fields that have vevs are $\varphi_5 =
v^\prime_1$, $\varphi_6 = v^\prime_2$ and $\varphi_3 = \alpha$. The last vev 
breaks charge conservation and would give mass to the photon. We can calculate 
the derivatives of the potential with respect to $v^\prime_1$, $v^\prime_2$ and 
$\alpha$ explicitly and arrive at the conclusion that there is always a solution
for $\alpha = 0$, which corresponds to the Normal extremum. It is however easier
to deal with the results if one uses the matrix notation. We define the vector 
$Y$ to have components $(x_1\,,\,x_2 \,,\,x_3\,,\,x_4)$ evaluated at the CB 
stationary point, that is,
\begin{equation}
Y\;\; = \;\; \begin{bmatrix} {v^\prime_1}^2 \\  {v^\prime_2}^2\,+\,\alpha^2 \\
v^\prime_1 v^\prime_2 \\ 0 \end{bmatrix} \;\;\;.
\label{eq:y}
\end{equation}
Unlike the Normal case we now have three independent variables, this allows us 
to write the stationarity conditions as
\begin{equation}
\left.\frac{\partial V}{\partial X} \right|_{X = Y} \;\; = \;\; 0 \;\; 
\Leftrightarrow \;\; A\; + \; B\,Y \;\; = \;\; 0 \;\;\;. 
\label{eq:ay}
\end{equation}
Hence, as long as the matrix $B$ is invertible, the solution $Y$ is such that
\begin{equation}
Y\;\; = \;\;-\,B^{-1}\, A\;\;\; .
\label{eq:vev}
\end{equation}
We observe that the CB vevs are given by a linear equation, which means that 
this solution, if it exists, is {\em unique}. This is unlike the Normal case, 
where the minimisation conditions~\eqref{eq:min} lead to a system of two coupled
cubic equations and as such can in principle produce multiple solutions. Notice 
however that the CB solution does not always exist even if $B$ is invertible - 
the first two components of the vector $Y$ must necessarily be positive, and not
all choices of $A$ and $B$ matrices will give such a result in 
eq.~\eqref{eq:vev}. Finally, the same reasoning that led us to 
eq.~\eqref{eq:vmin} can be applied to this stationary point and we may write the
value of the potential, $V_{CB}$, as
\begin{equation}
V_{CB} \;\; = \;\; \frac{1}{2}\,A^T\,Y \;\; = \;\; -\frac{1}{2}\,Y^T\, B\,Y 
\;\;.
\label{eq:vcb}
\end{equation}
We now have all the ingredients necessary to show that, if the Normal minimum 
exists, it is {\em always} deeper than the CB stationary point. We look again at
equation~\eqref{eq:vlm} and use equation~\eqref{eq:ay} to write
\begin{equation}
V^\prime \;\; = \;\; -\,B\, Y\,+\,B\, X_{N} \;\; .
\label{eq:vlp}
\end{equation}
Remembering that $X_{N}^T\,V^\prime \,=\,0$ we obtain
\begin{equation}
-\,X_{N}^T\,B\, Y \,+\,X_{N}^T\,B\, X_{N}\;\; = \;\; 0 
\end{equation}
and thus
\begin{equation}
X_{N}^T\,B\, Y \;\; = \;\; X_{N}^T\,B\, X_{N} \;\; = \;\; -\,2\,V_{N}
\;\; ,
\end{equation}
the last step arising from equation~\eqref{eq:vmin}. We now calculate the
quantity $Y^T\,V^\prime$,
\begin{equation}
Y^T\,V^\prime \;\; = \;\; -\,Y^T\,B\, Y  \,+\, Y^T\,B\, X_{N} \;\;\; .
\end{equation}
Because the matrix $B$ is symmetric we have $Y^T\,B\,X_{N} \,=\,X_{N}^T\,B\,Y
\,=\,-\,2\,V_{N}$ and, from equation~\eqref{eq:vcb}, $Y^T\,B\,Y \,=\, -\,2\,
V_{CB}$, so
\begin{equation}
V_{CB}\;-\;V_{N}\;\; = \;\; \frac{1}{2}\,Y^T\,V^\prime \;\;\; .
\end{equation}
The product $Y^T\,V^\prime$ can be explicitly written as 
(from~\eqref{eq:vl} and~\eqref{eq:y})
\begin{align}
Y^T\,V^\prime &= \;\; \left( -\frac{V^\prime_3}{2 v_1 v_2}\right)\;\left[
{v^\prime_1}^2\,v_2^2\;+\; ({v^\prime_2}^2\,+\,\alpha^2)\,v_1^2\;-\;2\,
v^\prime_1\,v^\prime_2\,v_1\,v_2 \right]\;\; \nonumber \\
 &= \;\;  \left( -\frac{V^\prime_3}{2 v_1 v_2}\right)\;\left[ (v^\prime_1\,v_2
\;-\;v^\prime_2\,v_1)^2\; + \; \alpha^2\,v_1^2\right]\;\;\; .
\end{align}
At the end of section~\ref{sec:sm} we have demonstrated that the quantity 
$-V^\prime_3/(2 v_1 v_2)$ is positive so we conclude that
\begin{equation}
V_{CB}\;-\;V_{N} \;\; > \;\; 0 \;\;\; ,
\end{equation}
which is to say, the Normal minimum is always deeper than the CB stationary 
point. So we conclude that the Normal minimum is perfectly stable and can never
tunnel to a charge-breaking vacuum. In fact, as we will see in 
section~\ref{sec:mass}, the matrix $B$ determines the character of the CB 
stationary point - if it is positive definite, so is the matrix of the squared 
masses at the CB stationary point. Multiplying both sides of eq.~\eqref{eq:vlp} 
by $B^{-1}$ first and then by ${V^\prime}^T$, it is simple to rewrite 
$Y^T\,V^\prime$ as 
\begin{equation}
Y^T\,V^\prime \;\; = \;\; -\,{V^\prime}^T\,B^{-1}\,V^\prime \;\;\;.
\end{equation} 
We have shown that when the Normal minimum exists ${V^\prime}^T\,B^{-1}\,
V^\prime$ is negative. This implies that the matrix $B^{-1}$ is not positive 
definite. So $B$ is also not positive definite. We will prove in the appendix
that its first entry, $2b_{11}$, is necessarily positive to prevent the 
potential from being unbounded from below. Hence $B$ cannot be negative definite
either. Therefore, the CB stationary point is necessarily a saddle point. 

\section{CP breaking}
\label{sec:cp}

Besides the Normal minimum or the CB stationary point, we have another possible
stationary point, one that spontaneously breaks CP conservation. In this case
the fields which have non-zero vevs are $\varphi_5 = v^\prime_1$, 
$\varphi_6 = v^\prime_2$ and $\varphi_7 = \delta$ - this last one breaks CP. 
The variables $x_i$, at this stationary point, are $x_1 = {v^\prime_1}^2\,+\,
\delta^2$, $x_2 = {v^\prime_2}^2$, $x_3 = v^\prime_1 v^\prime_2$ and $x_4 = -\,
v^\prime_2\, \delta$. We see that $x_4^2\,=\,x_1\,x_2\,-\,x_3^2$ and as such the
potential for this field configuration can be written as
\begin{align}
V \;\;=& a_1\, x_1\, + \,a_2\, x_2\, + \,a_3 x_3 \,+\nonumber \\
 & b_{11}\, x_1^2\, +\, b_{22}\, x_2^2\, +\, (b_{33}\,-\,b_{44})\, x_3^2\,
 +\, (b_{12}\,+\,b_{44})\, x_1 x_2\, +\, b_{13}\, x_1 x_3\, + \,b_{23}\,
 x_2 x_3 \;\; .
\end{align}
Defining the vector $Z$ and the square, symmetric matrix $B_{CP}$ to be
\begin{equation}
Z\;=\;\begin{bmatrix} {v^\prime_1}^2\,+\,\delta^2 \\ {v^\prime_2}^2 \\ 
v^\prime_1 v^\prime_2 \\ 0 \end{bmatrix} \;\;\; , \;\;\;
B_{CP}\;=\;\begin{bmatrix} 2 b_{11} & b_{12} + b_{44} & b_{13} & 0 \\ b_{12} + 
b_{44} & 2 b_{22} & b_{23} & 0 \\ b_{13} & b_{23} & 2 ( b_{33} - b_{44} ) & 0 \\
0 & 0 & 0 & 0 \end{bmatrix} \;\; ,
\end{equation}
we see that the potential at this stationary point may be written in terms of 
$\{v^\prime_1\,,\, v^\prime_2\,,\,\delta\}$ as 
\begin{equation}
V_{CP} \;=\; A^T\,Z \;+\; \frac{1}{2}\,Z^T \,B_{CP}\,Z \;\;\; .
\end{equation}
Then we are exactly in the conditions of the previous section: the CP 
breaking vevs are given by the equation
\begin{equation}
Z\;\;=\;\; - \,B_{CP}^{-1}\,A \;\;\; ,
\end{equation}
the value of the potential at this stationary point is given by
\begin{equation}
V_{CP} \;\; = \;\; \frac{1}{2}\,A^T\,Z \;\; = \;\; -\frac{1}{2}\,Z^T\, B_{CP}\,Z
\;\;,
\end{equation}
and repeating the steps of the previous section we find
\begin{equation}
V_{CP}\;-\;V_{N}\;\; = \;\; -\frac{1}{2}\,{V^\prime}^T\,B_{CP}^{-1}\,V^\prime
\;\; = \;\; \frac{1}{2}\,Z^T\,V^\prime \;\;\; .
\label{eq:cpb}
\end{equation}
Calculating $Z^T\,V^\prime$ explicitly we find
\begin{equation}
Z^T\,V^\prime\;\; = \;\;\left( -\frac{V^\prime_3}{2 v_1 v_2}\right)\;\left[ 
(v^\prime_1\,v_2 \;-\;v^\prime_2\,v_1)^2\; + \; \delta^2\,v_2^2\right]
\end{equation}
and so we conclude that
\begin{equation}
V_{CP}\;-\;V_{N}\;\; >\;\; 0\;\;\; .
\end{equation}
That is, once we have found a Normal minimum, it is always deeper than the 
CP-breaking stationary point. Again, we have proved that $B_{CP}$ is not
positive definite and, since its first entry is positive (see the appendix), it 
cannot be negative definite. As before $B_{CP}$ determines the nature of the 
squared mass matrix at this stationary point (see section~\ref{sec:mass}), which
means that if a Normal minimum exists, the CP stationary point is necessarily a 
saddle point. 

\section{Mass matrices}
\label{sec:mass}

To determine the nature of the stationary points one must analyse the second 
derivatives of the potential, which is to say, the scalar squared mass matrices.
They are given by
\begin{equation}
\frac{\partial^2 V}{\partial \varphi_i\partial \varphi_j}\;\; =\;\; \frac{
\partial V}{\partial x_l}\,\frac{\partial^2 x_l}{\partial \varphi_i\partial 
\varphi_j}\;+\;\frac{\partial^2 V}{\partial x_l\partial x_m}\,\frac{\partial 
x_l}{\partial \varphi_i}\,\frac{\partial x_m}{\partial \varphi_j} \;\;\; .
\label{eq:mm}
\end{equation}
Extending the notation of section~\ref{sec:sm} we define $V^\prime_i \,=\,
\partial V/\partial x_i$ evaluated at any of the three stationary points that 
we are considering. The first term in the above equation is written as an 
$8\times 8$ matrix of the form
\begin{equation}
[M^2_1]\;\; = \;\; \begin{bmatrix} M^2_{11} & 0 \\ 0 & M^2_{12} \end{bmatrix}
\end{equation}
where $M^2_{11}$ and $M^2_{12}$ are $4\times 4$ matrices given by
\begin{equation}
M^2_{11} \;=\; \begin{bmatrix} 2 V^\prime_1 & 0 & V^\prime_3 & V^\prime_4 \\ 0 &
2 V^\prime_1 & - V^\prime_4 & V^\prime_3 \\ V^\prime_3 & - V^\prime_4 & 2 
V^\prime_2 & 0 \\ V^\prime_4 & V^\prime_3 & 0 & 2 V^\prime_2 \end{bmatrix} \;\;,
\;\;M^2_{12} \;=\; \begin{bmatrix} 2 V^\prime_1 & V^\prime_3 & 0 & V^\prime_4 \\
V^\prime_3 & 2 V^\prime_2 & - V^\prime_4 & 0 \\ 0 & - V^\prime_4 & 2 V^\prime_1 
& V^\prime_3 \\ V^\prime_4 & 0 & V^\prime_3 & 2 V^\prime_2 \end{bmatrix} \;\; .
\end{equation}
In the second term of eq.~\eqref{eq:mm} the derivative $\partial^2 V/\partial 
x_l\partial x_m$ is clearly the matrix element $B_{lm}$ of the matrix $B$ 
defined in eq.~\eqref{eq:ab}. As for the derivatives $\partial x_i/\partial 
\varphi_j$, they form a $4\times 8$ matrix which we call $C$, given by
\begin{equation}
[C]\;\;=\;\; \begin{bmatrix} 0 & 0 & 0 & 0 & 2 \varphi_5 & 0 & 2\varphi_7 & 0 \\
0 & 0 & 2 \varphi_3 & 0 & 0 & 2 \varphi_6 & 0 & 0 \\ \varphi_3 & 0 &
0 & 0 & \varphi_6 & \varphi_5 & 0 & \varphi_7 \\ 0 & -\varphi_3 & 0 & 0 & 0 & 
-\varphi_7 & -\varphi_6 & \varphi_5 \end{bmatrix}  \;\;\; ,
\end{equation}
This matrix is evaluated at each of the different stationary points which is why
only the fields $\{\varphi_3\,,\,\varphi_5\,,\,\varphi_6\,,\,\varphi_7\}$ 
appear, the rest are always zero at the stationary points. Then the scalar 
squared mass matrix becomes
\begin{equation}
[M^2] \;\;=\;\; \frac{1}{2}\,\left([M_1^2]\;+\; C^T\,B\,C\right) \;\;\; ,
\label{eq:mmr}
\end{equation}
where the factor of 1/2 in the left hand-side is due to the fields $\varphi$ 
being real. Let us now analyse this matrix for each of the three cases. 

\begin{itemize} \item The Normal stationary point \end{itemize}

From section~\ref{sec:cb} we see that $V^\prime_4 = 0$ and ${V^\prime_3}^2 \,=\,
4\,V^\prime_1\,V^\prime_2$. In this case only $\varphi_5 = v_1$ and $\varphi_6 =
v_2$ are non zero, so that the first four columns of the matrix $C$ are zeros.
As a consequence the matrix $C^T\,B\,C$ is block diagonal, with the first 
diagonal $4\times 4$ block composed of zeros, that is,
\begin{equation}
C^T\,B\,C \;\;=\;\; \begin{bmatrix} 0 & 0 \\ 0 & B^\prime \end{bmatrix}\;\;\;,
\end{equation}
where $B^\prime$ is a $4\times 4$ matrix. In our notation this means that
the matrix $M^2_{11}/2$ is now the mass matrix of the charged Higgs, and it is
very simple to see that it has two zero eigenvalues and a doubly degenerate 
eigenvalue given by $V^\prime_1\,+\,V^\prime_2$. The matrices $M^2_{12}$ and
$B^\prime$ are also block-diagonal, with two $2\times 2$ non-zero matrices 
along the diagonal. One of these is the mass matrix of the pseudo-scalar sector,
given by
\begin{equation}
\begin{bmatrix} V^\prime_1\,+\,b_{44}\,v_2^2 & V^\prime_3\,-\,b_{44}\,v_1\,v_2 
\\ V^\prime_3\,-\,b_{44}\,v_1\,v_2 & V^\prime_2\,+\,b_{44}\,v_2^2 \end{bmatrix}
\;\;\;.
\end{equation}
This matrix has a zero eigenvalue and another one equal to $V^\prime_1\,+\,
V^\prime_2\,+\,b_{44}\,(v_1^2\,+\,v_2^2)$. The mass matrix for the CP-even 
scalar sector is given by
\begin{equation}
[M^2_{h,H}]\;\;=\;\; \begin{bmatrix} V^\prime_1\,+\,H_1 & V^\prime_3\,+\,H_3 \\ 
V^\prime_3\,+\,H_3 & V^\prime_2\,+\,H_2\end{bmatrix} \;\;\;,
\end{equation}
with
\begin{align}
H_1 &= 4\,b_{11}\,v_1^2\,+\,2\,b_{13}\,v_1\,v_2\,+\,b_{33}\,v_2^2 \nonumber \\
H_2 &= 4\,b_{22}\,v_2^2\,+\,2\,b_{23}\,v_1\,v_2\,+\,b_{33}\,v_1^2 \nonumber \\
H_3 &= (2\,b_{12}\,+\,b_{33})\,v_1\,v_2\,+\,\,b_{13}\,v_1^2\,+\,b_{23}\,v_2^2 
\;\;\;. \\
\end{align}

\begin{itemize} \item The CB stationary point \end{itemize}

The mass matrix in this case is very simple given that every $V^\prime_i$ is 
zero, which means $M^2_1 = 0$. The matrix $C$ now has only one column of zeros 
but it is very easy to see that three other columns are linearly dependent. With
judicious operations on the lines and columns of the matrix $C$ we manage to set
to zero its first four columns and therefore write 
\begin{equation}
[M^2] \;\;=\;\;\frac{1}{2}\,\begin{bmatrix} 0 & 0 \\ 0 & {C^\prime}^T\,B\,C^\prime
\end{bmatrix}
\end{equation}
where $C^\prime$ is a $4\times 4$ matrix. It is then obvious that $M^2$ has four
zero eigenvalues, exactly those we would expect having broken the $U(1)_{em}$
symmetry. The non-zero eigenvalues are then those of the matrix
\begin{equation}
[M^2_{CB}] \;\;=\;\;\frac{1}{2}\,{C^\prime}^T\,B\,C^\prime
\end{equation}
and this expression implies that $[M^2_{CB}]$ is positive definite if and only 
if $B$ is positive definite. If $B$ is negative or semi-positive definite
$[M^2_{CB}]$ is likewise defined. This justifies our assertion in 
section~\ref{sec:cb} that the CB stationary points are saddle points. 

With further operations on lines and columns it is still possible to simplify
$[M^2_{CB}]$: one of its eigenvalues is $b_{44}\,({v^\prime_1}^2\,+\,
{v^\prime_2}^2\,+\,\alpha^2)$, the remaining three are those of the following 
$3\times 3$ matrix,
\begin{equation}
\begin{bmatrix} CB_{11} & CB_{12} & CB_{13} \\ CB_{21} & CB_{22} & CB_{23} \\
CB_{31} & CB_{32} & CB_{33} \end{bmatrix} \;\;\; ,
\end{equation}
with
\begin{align}
CB_{11} &= (b_{13}\,+\,b_{23})\,v^\prime_1\,v^\prime_2\,+\,b_{33}\,
({v^\prime_1}^2 \,+\,{v^\prime_2}^2\,+\,\alpha^2) \nonumber \\
CB_{12} &= b_{33}\,v^\prime_1\,v^\prime_2\,+\,b_{23}\,({v^\prime_2}^2\,+\,
\alpha^2) \nonumber \\
CB_{13} &= b_{33}\,{v^\prime_2}^2\,+\,b_{13}\,v^\prime_1\,v^\prime_2 
\nonumber \\
CB_{21} &= b_{23}\,({v^\prime_1}^2\,+\,{v^\prime_2}^2\,+\,\alpha^2)\,+\,(4\,
b_{22} \,+\,2\,b_{12})\,v^\prime_1\,v^\prime_2 \nonumber \\
CB_{22} &= b_{23}\,v^\prime_1\,v^\prime_2\,+\,4\,b_{22}\,({v^\prime_2}^2\,+\,
\alpha^2) \nonumber \\
CB_{23} &= b_{23}\,{v^\prime_2}^2\,+\,2\,b_{12}\,v^\prime_1\,v^\prime_2
\nonumber \\
CB_{31} &= (4\,b_{11}\,+\,2\,b_{12})\,{v^\prime_1}^2\,+\,b_{13}\,
\frac{v^\prime_1}{v^\prime_2}\,({v^\prime_1}^2\,+\,{v^\prime_2}^2\,+\,\alpha^2) 
\nonumber \\
CB_{32} &= b_{13}\,{v^\prime_1}^2\,+\,2\,b_{12}\,\frac{v^\prime_1}{v^\prime_2}\,
({v^\prime_2}^2\,+\,\alpha^2) \nonumber \\
CB_{33} &= b_{13}\,v^\prime_1\,v^\prime_2\,+\,4\,b_{11}\,{v^\prime_1}^2 \;\; .
\end{align}

\begin{itemize} \item The CP stationary point \end{itemize}

Again the matrix $C$ has the first four columns equal to zero, which means that
the matrix $B$ only has an impact on the lower right $4\times 4$ corner of 
$[M^2]$. From the expressions of section~\ref{sec:cp} it is easy to obtain
$V^\prime_1\,=\,-\,b_{44}\,{v^\prime_2}^2$, $V^\prime_2\,=\,-\,b_{44}\,
({v^\prime_1}^2\,+\,\delta^2)$, $V^\prime_3\,=\,2\,b_{44}\,v^\prime_1\,
v^\prime_2$ and $V^\prime_4\,=\,-\,2\,b_{44}\,v^\prime_2\,\delta$. Once again 
the matrix $M^2_1$ is the squared mass matrix of the charged Higgs sector, it 
has two zero eigenvalues and a doubly degenerate non-zero one, given by $-\,
b_{44}\,({v^\prime_1}^2\,+\,{v^\prime_2}^2\,+\,\delta^2)$. The $\delta$ vev now 
causes mixing between the CP-even and odd scalar fields so that we are left with
a $4\times 4$ symmetric matrix, $M^2_{CP}$, whose entries are
\begin{align}
M^2_{CP}(1,1) &= 4\,b_{11}\,{v^\prime_1}^2\,+\,2\,b_{13}\,v^\prime_1\,v^\prime_2 \,+\,(b_{33}\,-\,b_{44})\,{v^\prime_2}^2 \nonumber \\
M^2_{CP}(1,2) &= b_{13}\,{v^\prime_1}^2\,+\,(b_{33}\,+\,b_{44}+\,2\,b_{12})\,
v^\prime_1\,v^\prime_2\,+\,b_{23}\,{v^\prime_2}^2 \nonumber \\
M^2_{CP}(1,3) &= (b_{13}\,v^\prime_2\,+\,4\,b_{11}\,v^\prime_1)\, \delta 
\nonumber \\
M^2_{CP}(1,4) &= [b_{13}\,v^\prime_1\,+\,(b_{33}\,-\,b_{44})\,v^\prime_2]\,
\delta \nonumber \\
M^2_{CP}(2,2) &= (b_{33}\,-\,b_{44})\,{v^\prime_1}^2\,+\,2\,b_{23}\,v^\prime_1\,
v^\prime_2 \,+\,4\,b_{22}\,{v^\prime_2}^2 \nonumber \\
M^2_{CP}(2,3) &= [b_{13}\,v^\prime_1\,+\,2\,(b_{12}\,+\,b_{44})\,v^\prime_2]\,
\delta \nonumber \\
M^2_{CP}(2,4) &= [b_{23}\,v^\prime_2\,+\,(b_{33}\,-\,b_{44})\,v^\prime_1]\,
\delta \nonumber \\
M^2_{CP}(3,3) &= 4\,b_{11}\,\delta^2 \nonumber \\
M^2_{CP}(3,4) &= b_{13}\,\delta^2 \nonumber \\
M^2_{CP}(4,4) &= (b_{33}\,-\,b_{44})\,\delta^2 \;\;\; .
\end{align}
This matrix has one zero eigenvalue which gives a total of three Goldstone 
bosons, as was to be expected. The most interesting aspect of the CP mass 
matrix lies in the value of the squared charged mass, here proportional to 
$-\,b_{44}$, whereas in the CB case it was proportional to $+\,b_{44}$. This 
means that we can never have, for the same choice of $b_{44}$, a CB minimum 
and a CP one, as was concluded in ref.~\cite{vel}. If we define the matrix 
$C^{\prime\prime}$ to be the restriction of matrix $C$ to its last four columns
it is trivial to show that
\begin{equation}
[M^2_{CP}] \;\;=\;\;\frac{1}{2}\,{C^{\prime\prime}}^T\,B_{CP}\,C^{\prime\prime}
\;\;\; .
\end{equation}
So, just like the CB case, the $B_{CP}$ matrix determines the nature of the
stationary point: if it is positive definite - and $b_{44}<0$, due to the 
charged Higgs eigenvalue - the stationary point is a minimum. If not, it is 
necessarily a saddle point.

\section{Conclusions}

In this paper we have shown a remarkable result: that the two Higgs doublet 
model is ``protected" against electric charge or CP spontaneous breaking. In 
other words, if the model has a minimum preserving $U(1)_{em}$ and CP, that
minimum is global. In this way, there is absolutely no possibility of tunneling
to deeper minima, and, for instance, the masslessness of the photon is 
guaranteed in these models. We also obtained a simple relation telling us that 
the only case where charge breaking might occur is when the matrix $B$ is 
positive definite. This situation implies that no Normal minima exists. For the 
future, when one is scanning the parameter space of the 2HDM one can safely 
exclude, from the beginning, the combinations of $b_{ij}$ parameters that give 
rise to a positive defined $B$. 

Charge breaking would be disastrous but there is considerable interest, from 
cosmologists to particle physicists, in models with the possibility of 
spontaneous CP violation. We have determined that this cannot happen for those 
ranges of parameters that lead to Normal minima. However, we have also 
established a very precise condition for spontaneous CP breaking to occur: CP is
spontaneously broken if and only ${V^\prime}^T\,B_{CP}^{-1}\,V^\prime > 
0$~\footnote{In reference~\cite{sil} CP violating quantities involving only the
Higgs sector were derived in models with explicit CP violation.}. In 
these circumstances the 2HDM no longer has a Normal minimum. A simpler condition
for spontaneous CP violation is to demand that the matrix $B_{CP}$ be positive 
definite - this condition, together with $b_{44}<0$, guarantees that the CP
stationary point, when it exists, is a minimum. We can also guarantee that if we
find a CP minimum, that too is safe against charge breaking. This arises from 
the fact that a CP minimum requires $b_{44}<0$ which automatically implies the 
matrix $B$ is {\em not} positive defined. Then we will have $V_{CB}\,-\,V_{N}\,
>\,0$ and $V_{CP}\,-\,V_{N}\,<\,0$. Therefore, $V_{CB}\,-\,V_{CP}\,>\,0$ and the
CP minimum is safe against charge breaking. 

Let us also stress that our conclusions are absolutely general, independent of 
particular values of the parameters of the theory, obviously. They hold for any 
of the more restricted models considered in ref.~\cite{vel}. It is simple to
recover the conditions presented in that reference to avoid CP minima by
analysing the matrix $B_{CP}$. We remark that the Higgs potential of the 
Supersymmetric Standard Model (SSM) is also included in the potentials we 
studied - in fact, it corresponds to the case $b_{11}\,=\, b_{22}\, =\,-\,
b_{12}/2\,=\,M_Z^2/(2 v^2)$, $b_{33}\,=\,b_{44}\,=\,2\,M_W^2/v^2$ and the 
remaining $b$ parameters set to zero, following the conventions of 
ref.~\cite{cast}. So we could conclude that at tree-level, the Supersymmetric
Higgs potential is safe against charge of CP violation, though this would not
preclude charge, colour or CP breaking arising from other scalar fields present
in those models. However, we must be cautious: it has been shown~\cite{gam} that
one-loop contributions to the minimisation of the potential have an enormous 
impact on charge breaking bounds in Supersymmetric models. Also, it was recently
shown~\cite{eu} that unless one performs a full one-loop calculation (both for 
the potential and the vevs, in both the CB potential and the ``normal" one) the
bounds one obtains can be overestimated. Therefore, we urge caution in applying
these conclusions to the SSM. Nevertheless one would expect the one-loop 
contributions to be much less important in the non-supersymmetric 2HDM due to
the much smaller particle content of the latter theory. 

\vspace{0.25cm}
{\bf Acknowledgments:} We are thankful to Pedro Freitas and Lu\'{\i}s Trabucho
for their assistance and discussions. This work is supported by Funda\c{c}\~ao 
para a Ci\^encia e Tecnologia under contract POCTI/FNU/49523/2002. P.M.F. is
supported by FCT under contract SFRH/BPD/5575/2001. 

\section*{Appendix: bounds on the $b$-parameters}

Requiring that the potential~\eqref{eq:pot} be bounded from below in all 
possible directions imposes some interesting bounds on the $b$ parameters. These
bounds are obtained studying particular field directions and requiring that the 
quartic terms' limit as the fields go to infinity be positive or at most zero. 
For instance, we can choose a direction such that all variables $x$ except $x_1$
are zero, and $x_1 \rightarrow \infty$ (for instance, by choosing $\varphi_1 
\rightarrow\infty$ and all remaining $\varphi_i$ equal to zero). Then along this
direction the potential goes to infinity as $b_{11}\,\varphi_1^4$, and if we
want this limit to be positive or zero, we must demand that $b_{11} \geq 0$. 
Likewise, by choosing $\varphi_3 \rightarrow \infty$ and all others zero, for 
instance, we would obtain the condition $b_{22} \geq 0$. We cannot obtain a 
similar condition for $b_{33}$ and $b_{44}$ because we can never find a field 
direction with $x_{3,4} \rightarrow\infty$ without having $x_{1,2}$ diverging as
well. Let us now consider the direction $\varphi_1,\varphi_5 \rightarrow \infty$
and all others set to zero. Only $x_1$ and $x_2$ are non-zero and the potential
is reduced to the terms $b_{11}\,\varphi_1^4\,+\,b_{22}\,\varphi_5^4\,+\,b_{12}
\,\varphi_1^2\,\varphi_5^2$. By choosing polar coordinates such that 
$\varphi_1^2\,=\,r\,\cos\theta$ and $\varphi_2^2\,=\,r\,\sin\theta$, the 
potential will always be greater or equal to zero at infinity along this 
direction if, for any $0 \leq \theta \leq \pi/2$, 
\begin{equation}
b_{11}\,\cos^2\theta\,+\,b_{22}\,\sin^2\theta\,+\,b_{12}\,\sin\theta\cos
\theta\;\; \geq \;\; 0 \;\;\;.
\end{equation}
A simple minimisation in $\theta$ shows that this occurs as long as
\begin{equation}
b_{12}\;\;\geq \;\; -\,2\-\sqrt{b_{11}\,b_{22}} \;\;\; .
\label{eq:b12}
\end{equation}
Likewise, if we choose the direction $\varphi_1, \varphi_4 \rightarrow \infty$ 
and again use polar coordinates, the boundedness from below condition translates
into
\begin{equation}
b_{11}\,\cos^2\theta\,+\,b_{22}\,\sin^2\theta\,+\,(b_{12}\,+\,b_{44})\sin\theta
\cos \theta\;\; \geq \;\; 0 
\end{equation}
and so the condition we obtain is similar to the previous one, 
\begin{equation}
b_{12}\,+\,b_{44}\;\;\geq \;\; -\,2\-\sqrt{b_{11}\,b_{22}} \;\;\; .
\label{eq:b44}
\end{equation}
With the direction $\varphi_1, \varphi_3 \rightarrow \infty$ we can study what
happens with the $b_{3i}$ parameters, the condition we obtain is (making 
$\varphi_1 \,=\,r\,\cos\theta$ and $\varphi_3\,=\,r\,\sin\theta$ so that now 
there are no restrictions on the value of $\theta$)
\begin{equation}
b_{11}\cos^4\theta\,+\,b_{22}\sin^4\theta\,+\,(b_{12}+b_{33})\sin^2
\theta\cos^2\theta\,+\,b_{13}\cos^3\theta\sin\theta\,+\,b_{23}\sin^3\theta
\cos \theta\; \geq \; 0 \;\; .
\end{equation}
Since this condition has to hold for any value of $\theta$ we can make the 
change $\theta\,\rightarrow \,-\,\theta$ and obtain
\begin{equation}
|b_{13}\,\cos^3\theta\sin\theta\,+\,b_{23}\,\sin^3\theta\cos\theta| \,\leq\,
b_{11}\cos^4\theta\,+\,b_{22}\sin^4\theta\,+\,(b_{12}+b_{33})\sin^2
\theta\cos^2\theta\;\; .
\label{eq:3}
\end{equation}
So, we conclude that, for any $\theta$, we have
\begin{equation}
(b_{11}\,\cos^4\theta\,+\,b_{22}\,\sin^4\theta\,+\,(b_{12}\,+\,b_{33})\sin^2
\theta\cos^2\theta \;\;\geq \;\; 0 
\end{equation}
and we obtain a condition similar to~\eqref{eq:b12} and~\eqref{eq:b44},
\begin{equation}
b_{12}\,+\,b_{33}\;\;\geq \;\; -\,2\-\sqrt{b_{11}\,b_{22}} \;\;\; .
\end{equation}
Finally, making $\theta = \pi/4$ in eq.~\eqref{eq:3}, we obtain a more 
manageable bound on $b_{13}$ and $b_{23}$, 
\begin{equation}
|b_{13}\,+\,b_{23}|\;\; \leq \;\; b_{11}\,+\,b_{22}\,+\,b_{12}\,+\,b_{33}\;\;\;.
\end{equation}


\begin{thebibliography}{99}
\bibitem{lee} T.D. Lee, {\em Phys. Rev.} {\bf D8} (1973) 1226;

G.C. Branco and M. N. Rebelo, {\em Phys. Rev.} {\bf D22} (1980) 2901.
\bibitem{err} L.Mclerron {\em et al}, {\em Phys. Lett.} {\bf B256} (1991) 451. 
\bibitem{sher} M. Sher, {\em Phys. Rep.} {\bf 179} (1989) 273; 

G.C. Branco, L. Lavoura and J.P. Silva, {\em CP Violation} (Oxford University
Press, Oxford, England, 1999). 
\bibitem{fre} J.M. Fr\'ere, D.R.T. Jones and S. Raby,
{\em Nucl. Phys.} {\bf B222} (1983) 11.
\bibitem{eve} L. Alvarez-Gaum\'e, J. Polchinski and M. Wise, {\em Nucl. Phys.}
{\bf B221} (1983) 495; 

J.P. Derendinger and C.A. Savoy {\em Nucl. Phys.} {\bf B237} 307; 

C. Kounnas, A.B. Lahanas, D.V. Nanopoulos and M. Quir\'os, {\em Nucl. Phys.} 
{\bf B236} (1984) 438; 

M. Claudson, L.J. Hall and I. Hinchliffe, {\em Nucl. Phys.} {\bf B228} (1983) 
501; 

M. Drees, M. Gl\"uck and K. Grassie, {\em Phys. Lett.} {\bf B157} (1985) 164;

J.F. Gunion, H.E. Haber and M. Sher, {\em Nucl. Phys.} {\bf B331} (1988) 320.
\bibitem{phe} U. Ellwanger and C. Hugonie, {\bf hep-ph/9811386}; 

S. Abel and T. Falk, {\em Phys. Lett.} {\bf B444} (1998) 427; 

S. Abel and C. Savoy, {\em Phys. Lett.} {\bf B444} (1998) 119; 

S. Abel and B. Allanach, {\em Phys. Lett.} {\bf B431} (1998) 339.

OPAL Collaboration, {\em Eur. Phys. Jour} {\bf C7} (1999) 407; {\em ibid.}, 
{\em Eur. Phys. Jour} {\bf C12} (2000) 567.
\bibitem{haber} J.F. Gunion and H.E. Haber, {\em Phys. Rev.} {\bf D67} (2003)
075019.
\bibitem{vel} J. Velhinho, R. Santos e A. Barroso, {\em Phys. Lett.} {\bf B322}
(1994) 213.
\bibitem{sil} L. Lavoura dn J.P. Silva, {\em Phys. Rev.} {\bf D50} (1994) 4619.
\bibitem{cast} D.J. Casta\~no, E.J. Piard and P. Ramond, {\em Phys. Rev.} 
{\bf D49} (1994) 4882.
\bibitem{gam} G. Gamberini, G. Ridolfi and F. Zwirner, {\em Nucl. Phys.} {\bf
B331} (1990) 331.
\bibitem{eu} P.M. Ferreira, {\em Phys. Lett.} {\bf B509} (2001) 120;

P.M. Ferreira, {\em Phys. Lett.} {\bf B512} (2001) 379. 
\end{thebibliography}
\end{document}